\documentclass[12pt]{iopart}

\usepackage{iopams}

\expandafter\let\csname equation*\endcsname\relax

\expandafter\let\csname endequation*\endcsname\relax

\usepackage{amsmath,amssymb,amsbsy,epsfig,color,graphicx}
\usepackage[ansinew]{inputenc}
\usepackage[english]{babel}
\usepackage{braket}
\usepackage{ulem}
\usepackage{dcolumn}
\usepackage{hyperref}

\newcommand{\bea}{\begin{eqnarray}}
\newcommand{\eea}{\end{eqnarray}}
\newcommand{\be}{\begin{equation}}
\newcommand{\ee}{\end{equation}}

\begin{document}

\title{The Frustration of being Odd: How Boundary Conditions can destroy Local Order}

\author{Vanja Mari\'{c}$^{1,2}$, Salvatore Marco Giampaolo$^{1}$, Domagoj Kui\'{c}$^{1}$, Fabio Franchini$^{1}$}

\address{$^{1}$ Division of Theoretical Physics, Ru\dj{}er Bo\v{s}kovi\'{c} Institute, Bijeni\u{c}ka cesta 54, 10000 Zagreb, Croatia}
\address{$^{2}$ SISSA and INFN, via Bonomea 265, 34136 Trieste, Italy.}

\begin{abstract}
A central tenant in the classification of phases is that boundary conditions cannot affect the bulk properties of a system. 
In this work, we show striking, yet puzzling, evidence of a clear violation of this assumption.
We use the prototypical example of an XYZ chain with no external field in a ring geometry with an odd number of sites and both ferromagnetic and antiferromagnetic interactions. 
In such a setting, even at finite sizes, we are able to calculate directly the spontaneous magnetizations that are traditionally used as order parameters to characterize the system's phases.
When ferromagnetic interactions dominate, we recover magnetizations that in the thermodynamic limit lose any knowledge about the boundary conditions and are in complete agreement with standard expectations.
On the contrary, when the system is governed by antiferromagnetic interactions, the magnetizations decay algebraically to zero with the system size and are not staggered, despite the AFM coupling. 
We term this behavior {\it ferromagnetic mesoscopic magnetization}.
Hence, in the antiferromagnetic regime, our results show an unexpected dependence of a local, one--spin expectation values on the boundary conditions, which is in contrast with predictions from the general theory.
\end{abstract}

\maketitle

\section{Introduction} 

Landau theory is one of the most impactful constructions of the last century.
It allows distinguishing between different phases through different local order parameters, quantities which are finite or vanish depending on the phase of a system~\cite{LandauStatPhys,AndersonNotions,ChandlerBook,ColemanBook}.
Although the new century has taught us that this classification is not complete, because certain phases of quantum matter are characterized by non--local order (for instance, topological~\cite{StoneBook,WenBook,Nayak2008,Hasan2010,FradkinBook,BernevigBook,Giampaolo15,Witten16,ZengBook}), Landau theory remains a cornerstone to interpret phases, directly borrowed from classical statistical mechanics.

Order parameters are supposed to capture macroscopic properties of phases and thus are believed not to depend on boundary conditions.
Indeed, as the boundary contributions are typically sub--extensive, they should bring a negligible effect for sufficiently large systems. 
Of course, depending on the system and on the type of interactions, there can be ambiguities on what ``sufficiently large system'' means, as sometimes boundary effects can decay just algebraically, even in phases considered gapped~\cite{Dong16,Giampaolo18}. 
Thus, the standard prescription to characterize a phase is to take the thermodynamic limit before evaluating physical observables~\cite{LandauStatPhys,AndersonNotions}.

This being said, the effects of boundary conditions have been a subject of interest in different contexts. 
For instance, the Kondo Effect can largely be interpreted as a boundary effect~\cite{Affleck95,Durganandini96}. 
But additional simple examples that have received a lot of attention immediately come to mind, such as conformal field theories (CFTs) and integrable models. 
In the former, conformal invariance poses tight bounds on the bulk properties and it has been established that boundary condition modifies the system's equilibrium behavior~\cite{Cardy84,Cardy04,DiFrancescoBook}. 
In the latter, different boundary conditions are commonly employed to study various properties.
For instance, the partition functions of 2D classical systems with domain wall boundary conditions provide the normalization of the corresponding quantum wave--functions~\cite{KorepinBook}. 
But certain boundary conditions are also known to generate the phenomenon of the {\it``arctic curve''}, which separate frozen regions (due to boundary conditions) from liquid ones~\cite{Korepin00,ZinnJustin02,Colomo10,BleherBook,Colomo16,Allegra16,Reshetikhin17,DiFrancesco18,Colomo18}.

A particularly thorny issue is represented by {\it frustration}~\cite{Toulouse77,Vannimenus77}. 
This term evokes different phenomena to different ears. 
While it simply refers to the presence of interactions promoting incompatible orderings (hence the impossibility of simultaneously minimizing every term in the system's Hamiltonian~\cite{Wolf03,Giampaolo11,Marzolino13}), the effects of frustration are varied and complex~\cite{Sadoc07,Lacroix11,Diep13}. 
Frustrated systems are a debated and very active field of research, with a rich phenomenology (different in many ways from that of non-frustrated systems) and with unique challenges~\cite{Lacroix11}. 
Nonetheless, at the heart of every frustrated system one can find one (or typically many) frustrated loops, which are the building blocks out of which the different phenomenologies arise~\cite{Wannier50}. 
Here, we will concentrate on this simplest, and original, incarnation of {\it geometrical frustration}. 
This is a classical concept which applies, for instance, to Ising spins coupled anti--ferromagnetically. 
While, locally, there is no problem in satisfying the AFM interactions, when the spins are arranged in a loop made of an odd number of sites, at least one bond needs to display ferromagnetic alignment. 
In this case, the frustration arises because of an incompatibility between local interactions and the global structure of the system and is due to the particular choice of boundary conditions (namely, periodic with an odd number of sites, which we term ``{\it frustrated boundary conditions}'', FBC). 
It is worth to note that, while for loops with an even number of sites the lowest energy state is doubly degenerate (given by the two types of Neel orders), with frustration the degeneracy becomes extensive because the defect can be placed on any bond of the ring.
Hence, by adding a single site to a ring, the problem turns from a non-degenerate perturbative one to a problem of degenerate perturbation theory, which is a priori non-trivial due to the extensive degeneracy in the thermodynamic limit.

Upon adding quantum interactions to a geometrically frustrated system, we can generally expect the degeneracy to be lifted. 
A perturbative approach characterizes the resulting ground state as the superposition of a delocalized excitation on top of the non frustrated ground state. 
This picture was recently checked in~\cite{Giampaolo18} for its validity even beyond the perturbative regime and confirmed using the Entanglement Entropy. 
The effects of such delocalized excitation have been studied in the past~\cite{Burkhards1985,Cabrera1987,Campostrini15}, revealing subtle phenomena, usually dismissed by the community as peculiar quirks, because these analyses never addressed local observables, but rather properties such as the spectral gap.

In this work, we pluck a hole in this canvas by focusing instead on the order parameter of antiferromagnetic spin chains and by showing that FBC makes it vanish as the ring's length diverges.
To the best of our knowledge, this is the first example of a case in which boundary conditions affect local observables (in thermodynamic limit) and it is in evident contrast with standard general arguments recapped above.
Nonetheless, we should remark that evidence pointing toward a vanishing of the spontaneous magnetization with FBC was already reported, for instance through the two--point function~\cite{Dong16,Giampaolo18}, but in previous works, the importance of finite--size effects was harder to estimate. 
We should also stress that our result is consistent with the single--particle picture mentioned above, as the traveling excitation destroys local order by flipping every spin in its motion.
At the moment, we do not know how to reconcile our findings with the traditional paradigm, although we can speculate that, being geometrical frustration a non--local effect, some sort of topological mechanism is at play so that we propose to call the interplay between quantum interaction and FBC with the term ``{\it topological frustration}''. 
Our evidence is inescapable and should compel the community to understand what makes the spin chains we consider different and so sensitive to the boundary conditions, so to understand what class of models share the same or similar behaviors and how do they fit in the standard paradigm.

After introducing the systems under consideration (namely, a class of spin--$\frac{1}{2}$ chains with a global $\mathbb{Z}_2$ symmetry), we will recap the two complementary approaches to extract the order parameter in the symmetry broken phase in absence of frustration and then apply the same techniques to the case with FBC. 
Doing so, we benchmark our technique, showing that it yields the expected results in the former case, while it shows that frustration suppresses the order parameter to zero. 
In our work, we consider both models where we can perform exact analytical calculations to prove our claims and generalizations in which we have to resort to numerical diagonalization. 
We will show that for non--frustrated systems the order parameter grows to saturation exponentially with growing total system size, while it decreases toward zero algebraically with FBC.

\section{The spin chains and their properties}

We consider an anisotropic spin--$\frac{1}{2}$ chain with Hamiltonian 
\bea
  H & = & \sum\limits_{j=1}^N 
      \cos \delta \left(\cos\phi \, \sigma_j^x\sigma_{j+1}^x  
      + \sin\phi \, \sigma_j^y\sigma_{j+1}^y \right)  - \sin \delta   \sigma_j^z\sigma_{j+1}^z \, ,
  \label{Hamiltonian}
\eea
where $\sigma_j^\alpha$, with $\alpha=x,y,z$, are Pauli operators and $N$ is the number of lattice sites, which we henceforth set to be odd $N=2M+1$. 
Crucially, we apply periodic boundary conditions $\sigma_{j+N}^\alpha = \sigma_j^\alpha$.

The model is expected to exhibit a quantum phase transition every time two of the couplings are, in magnitude, equal and greater than the third~\cite{Ercolessi13} (in that case, the model becomes equivalent to a critical XXZ chain~\cite{Franchini17}). 
Dualities are connecting different rearrangements of the couplings along the $x$, $y$, and $z$ directions~\cite{Ercolessi13}. 
Moreover, to avoid additional effects (and degeneracies) that will be the subject of subsequent works~\cite{Maric19-1,Maric19-2}, we will allow only one antiferromagnetic (AFM) coupling (namely, along the $x$ direction), letting the other two to favor a ferromagnetic alignment. 
We thus limit the range of the anisotropy parameters such that $\phi\in[-\pi/2,0]$ and $\delta\in[0,\pi/2]$, so that the phase transition is at $\phi=-\pi/4$ (for $\tan \delta < 1/\sqrt{2}$) and separates two phases characterized by a two--fold degenerate ground state.
In particular, for $\phi \in \left[-\pi/2,-\pi/4 \right)$ the phase favors a ferromagnetic alignment along the $y$ direction (yFM), while for $\phi \in \left(-\pi/4, 0 \right]$ the dominant interaction is AFM along the $x$ direction (xAFM) and thus {\it topologically frustrated}.

With no external field, all three parity operators along the three axes $\Pi^\alpha \equiv \bigotimes_{j=1}^{N} \sigma_j^\alpha$ commute with the XYZ Hamiltonian in eq.~(\ref{Hamiltonian}) ($\left[H,\Pi^\alpha \right]$).
Moreover, since we are considering systems made by an odd number of sites $N=2M+1$, the $\Pi^\alpha$ do not commute with one another, but rather anti--commute $\left\{ \Pi^\alpha, \Pi^\beta \right\} = 2 \delta_{\alpha,\beta}$ and actually fulfill a $SU(2)$ algebra.
This structure implies that every state is exactly degenerate an even number of times, also on a finite chain. 
If $\ket{\Psi}$ is an eigenstate, say, of $\Pi^z$, then $\Pi^x \ket{\Psi}$, that differs from $\Pi^y \ket{\Psi}$ by a global phase factor, is also an eigenstate of the Hamiltonian with opposite $z$--parity but with the same energy.

Applying an external magnetic field $h$ along, say, the $z$--direction leaves only $\Pi^z$ to commute with the Hamiltonian, thus restoring the original $\mathbb{Z}_2$ symmetry the model is known for and breaking the exact finite-size degeneracy between the states~\cite{Franchini17, Sachdev11}.
Nonetheless, up to a critical value of $h$, it is known that the induced energy split is exponentially small in the system size~\cite{damski12} and thus that the degeneracy is restored in the thermodynamic limit, representing one of the simplest, and most cited, examples of spontaneous symmetry breaking (SSB)~\cite{Sachdev11}.
To simplify things, let us set $\delta=0$, so that eq.~(\ref{Hamiltonian}) describes an anisotropic XY chain~\cite{Franchini17,LSM-1961}. 
For $|h|<1$ we are in the SSB phase. 
This means that, although a ground state with definite \mbox{$z$--parities} necessarily has zero expectation value concerning $\sigma_j^x$ and $\sigma_j^y$, in the thermodynamic limit the degeneracy allows to select a ground state which is a superposition of different \mbox{$z$--parities}, which can thus display a spontaneous magnetization in the $x$ or $y$ direction.
In the yFM phase we expect the order parameter $m_y \equiv \braket{\sigma_j^y}$ to be finite, while in the xAFM the non--vanishing order parameter should be the staggered magnetization $m_x \equiv (-1)^j\braket{\sigma_j^x}$. 

\section{The ferromagnetic case}

Let us now turn back to the system in eq.~(\ref{Hamiltonian}) and focus on the ferromagnetic region \mbox{$\phi \in \left[-\pi/2,-\pi/4\right)$}.
The (quasi--)long--range order represented by the order parameter can be extracted in two ways: either from the two--point function or by selecting a suitable superposition of states at finite sizes and then following their magnetization toward the thermodynamic limit.
While the second is easily amenable to numerical approaches, the former, which is most suitable for analytical techniques, takes advantage of the cluster decomposition property
\begin{equation}
\label{cluster_decomposition}
\lim_{r\to\infty}\ \braket{\sigma_j^\alpha \sigma_{j+r}^\alpha}-\braket{\sigma_j^\alpha}\braket{\sigma_{j+r}^\alpha} =0 ,
\end{equation}
to extract the order parameter from the large distance behavior of the system's two--point correlators.
 
Exploiting the Jordan-Wigner Transformation~\cite{Jordan28,Franchini17}, which maps the spin degrees of freedom into spin-less fermions:
\begin{equation}
c_j=\Big(\bigotimes_{l=1}^{j-1}\sigma_l^z\Big)\frac{\sigma_j^x+\imath \sigma_j^y}{2} \; , \quad c_j^\dagger=\Big(\bigotimes_{l=1}^{j-1}\sigma_l^z\Big)\frac{\sigma_j^x-\imath \sigma_j^y}{2} \; ,
\end{equation}
the XY model can be brought into a free fermionic form. Before doing so, however, states must be separated according to their parity $\Pi^z$, since negative (positive) parity corresponds to (anti-)periodic boundary conditions applied to the fermions. Thus, the XY chain Hamiltonian can be written as 
\begin{equation}\label{H parity decomposition}
H=\frac{1+\Pi^z}{2}H^+ \frac{1+\Pi^z}{2} + \frac{1-\Pi^z}{2}H^- \frac{1-\Pi^z}{2} \; ,
\end{equation} 
where the exact expressions of $H^+$ and $H^-$ can be found in the appendix.
From these Hamiltonians it is possible to determine, following the method described in details in the appendix, the fundamental two--point correlation functions. 
These correlations are expressed as the determinant of a Toeplitz matrix, whose asymptotic behavior can be evaluated analytically~\cite{McCoy68}:
\begin{eqnarray}
\!\!\!\!\!\!\!\!\!\!\!\!\!\!\!\!\!\!\!\!\!\!\!\!\! \label{xx-yFM} \braket{\sigma_j^x \sigma_{j+r}^x} &\stackrel{r \to \infty}{\simeq} &\frac{2}{\pi\sqrt{1-\cot^2\phi}}\frac{\cot^r\phi}{r}+\ldots\\ 
\!\!\!\!\!\!\!\!\!\!\!\!\!\!\!\!\!\!\!\!\!\!\!\!\! \label{yy-yFM} \braket{\sigma_j^y \sigma_{j+r}^y} &\stackrel{r \to \infty}{\simeq} &
 \left\{
 \begin{array}{ll}
  \sqrt{1-\cot ^2\phi}\left(1+\frac{4}{\pi}\Big(\frac{\cot\phi}{1-\cot^2\phi}\Big)^2 \frac{\cot^r\phi}{r^2}+\ldots\right) & r=2m\\
  \sqrt{1-\cot^2\phi}\left(1+\frac{2}{\pi}\Big(\frac{\cot \phi}{1-\cot^2\phi}\Big)^2\frac{1+\cot^2\phi}{\cot\phi}\frac{\cot^r\phi}{r^2}+\ldots\right)& r=2m+1
 \end{array}
 \right.\\
\!\!\!\!\!\!\!\!\!\!\!\!\!\!\!\!\!\!\!\!\!\!\!\!\!  \label{zz-yFM} \braket{\sigma_j^z \sigma_{j+r}^z} & \stackrel{r\to\infty}{\simeq} &
  \left\{
 \begin{array}{ll}
  0 & r=2m\\
  -\frac{2}{\pi} \frac{\cot^r\phi}{r^2}+\ldots & r=2m+1
 \end{array}
 \right.
\end{eqnarray}

From these large $r$ behavior, taking into account the cluster decomposition hypothesis, we can extract the different magnetizations $ m_\alpha \equiv \braket{\sigma_j^\alpha}$
\bea
  m_x = m_z = 0,  \qquad 
  m_y = \left( 1 - \cot^2 \phi \right)^{1/4} \; .
  \label{yFMOrdParam}
\eea

However, on an odd--length chain at $h=0$, exploiting the symmetries that we have already illustrated, we can provide a direct way to evaluate the different magnetizations even in finite systems. 
In fact, if $\ket{g_z}$ is one of the degenerate ground states with definite $z$--parity which can be constructed in terms of the Bogoliuobov fermions~\cite{Franchini17}, we can generate a ground state with definite $x$--parity ($y$--parity) as $\ket{g_x} \equiv \frac{1}{\sqrt{2}}\left(1 + \Pi^x \right) \ket{g_z}$, ($\ket{g_y} \equiv \frac{1}{\sqrt{2}}\left(1 + \Pi^y \right) \ket{g_z}$).
All these states have a vanishing magnetization in the orthogonal directions while along their own axes we have
\bea
 \bra{g_x} \sigma_j^x \ket{g_x} = &
 \bra{g_z} \sigma_j^x \Pi^x \ket{g_z} = &
 \bra{g_z} \tilde{\Pi}_j^x \ket{g_z} \nonumber \\
 \bra{g_y} \sigma_j^y \ket{g_y} = &
 \bra{g_z} \sigma_j^y \Pi^y \ket{g_z} = &
 \bra{g_z} \tilde{\Pi}_j^y \ket{g_z} 
\label{trick}
\eea
where $\tilde{\Pi}_j^{\alpha} \equiv \bigotimes_{l \neq j} \sigma^{\alpha}_l$ with $\alpha$ running between $x$ and $y$. 
These states are the analytical continuation at \mbox{$h=0$} of the zero--temperature ``thermal'' ground state that spontaneously breaks the $\mathbb{Z}_2$ symmetry.

Note that in this way, we turn the calculation of the expectation value of an operator defined on a single--spin with respect to a ground state with a mixed $z$--parity into that of a string made by an even number of spin operators on a definite $z$--parity state, which is a standard problem.
Thus the RHS of eq.~(\ref{trick}) can be written again as the determinant of a Toeplitz matrix, whose asymptotic behavior can be studied analytically, similarly to what has been done in~\cite{Barouch71}. 
This novel ``trick'' can be understood as originating from the fact that, at zero external fields, the chain eq.~(\ref{Hamiltonian}) has particle/hole duality and that, on a chain with an odd number of sites, this symmetry relates states with different parities. 
The result of such analysis reproduces eq.~(\ref{yFMOrdParam}), proving the consistency of the two methods of evaluation for the order parameters.
More details on this direct approach in the appendix. 

\begin{figure}
\begin{center}
    \includegraphics[width=.8\columnwidth]{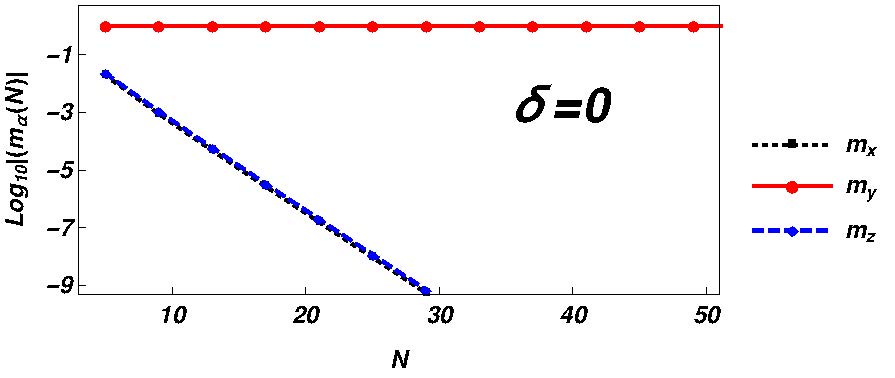}
    \includegraphics[width=.8\columnwidth]{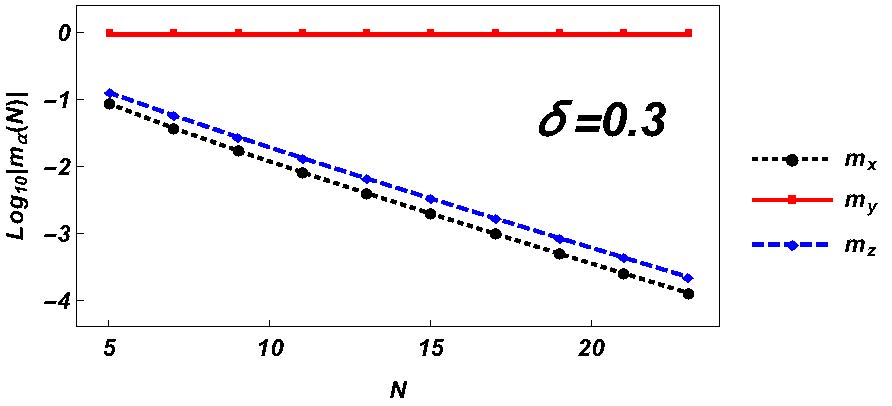}
\end{center}
    \caption{
    Magnetizations along the three axis (in absolute value) as a function of the chain length for the yFM phase at $\phi=-1.32$. 
    The upper panel dots represent the data gathered settings $\delta=0$ and using the ``trick'' discussed in the text to evaluate the magnetizations as determinants of $\frac{N-1}{2}\times \frac{N-1}{2}$ matrices, while the dots in the lower one are obtained taking \mbox{$\delta =.3$} and using exact numerical diagonalization.
    Regardless the value of $\delta$, $m_y$ quickly saturates to its asymptotic finite value, while $m_x$ and $m_z$ decay to zero exponentially fast, as shown by the best fit lines (plots presented in logarithmic scale).}
    \label{fig:FerroN}
\end{figure}

While for $\delta=0$ we can evaluate the magnetizations using the analytical ``trick'', for $\delta \ne 0$ we have to resort to numerical solutions.
In Fig.~\ref{fig:FerroN} we present some typical results for the finite size magnetizations for the XY and XYZ chain, showing a quick exponential decay in $N$ of $m_x$ and $m_z$ to zero and a fast saturation of $m_y$ (note that each plotted magnetization $m_\alpha$ is calculated with respect to the corresponding ground state $\ket{g_\alpha}$).

\section{The frustrated case}

We now turn to the case with ($\phi \in \left(-\pi/4, 0 \right]$), where the boundary conditions induce topological frustration.  
The effect of frustration has been recently studied in detail in Refs.~\cite{Dong16,Giampaolo18,Dong17}. 
For $\delta=0$, the model can be solved through the same steps used in the traditional cases and exactly mapped into a system of free fermions. 
In the ferromagnetic phase, the degeneracy between the different parity states is due to the presence of a single negative energy mode (only in one of the parity sectors), whose occupation lowers the energy of those states. 
With frustrations, the negative energy mode moves into the other parity sector and, because of the parity selection rules in \eqref{H parity decomposition}, it cannot be excited alone. 
Therefore, the effect of frustration is that the lowest energy states in each parity sector in \eqref{H parity decomposition} are not admissible.

The two degenerate ground states thus carry the signature of a single delocalized excitation and lie at the bottom of a band of states in which this excitation moves with different momenta (with an approximate Galilean dispersion relation).
Hence, another effect of frustration is to close the gap that would otherwise exist. 

Let us then repeat the extraction of the order parameters in the xAFM phase, following the same procedure we followed for yFM.
However, in the present case, the analytical computation of the spin correlations along the $x$ and $y$ directions requires the knowledge of the asymptotic behavior of a new type determinants, whose symbol contains a delta function with a peak at the momentum of the excitation. 
The details of the analytical computation are given in~\cite{Maric-math}, where such determinants are studied. 
Here we check these results numerically. The details are given in the appendix, where also the correlation along $z$ is computed. The results are
\begin{eqnarray}
\!\!\!\!\!\!\!\!\!\!\!\!\!\!\!\!\!\!\!\!\!\!\!\!\!\!\!\!\!\!\!\!\!\!\!\!
\label{xx-xAFM} \braket{\sigma_j^x \sigma_{j+r}^x}\! &\! 
 \stackrel{r \to \infty}{\simeq} \! & \!
 \left\{
 \begin{array}{ll}
  \!\!\sqrt{1-\tan^2\phi}\ \Big(1-\frac{2r}{N}\Big) \bigg[1+\frac{4}{\pi}\bigg(\frac{\tan \phi}{1-\tan^2\phi}\bigg)^2\frac{\tan^r\phi}{r^2}+\ldots\bigg] & \!\!r\!=\!2m \\
   \!\!-\sqrt{1-\tan^2\phi}\Big(1-\frac{2r}{N}\Big) \bigg[1+\frac{2}{\pi}\bigg(\frac{\tan \phi}{1-\tan^2\phi}\bigg)^2 \frac{1+\tan^2\phi}{\tan \phi} \frac{\tan^r\phi}{r^2}+\ldots\bigg] &\!\! r\!=\!2m\!+\!1
 \end{array}
 \right. \\
 \!\!\!\!\!\!\!\!\!\!\!\!\!\!\!\!\!\!\!\!\!\!\!\!\!\!\!\!\!\!\!\!\!\!\!\!
\label{yy-xAFM} \braket{\sigma_j^y \sigma_{j+r}^y} \! &\! 
 \stackrel{r \to \infty}{\simeq} \! & \!
\left\{
\begin{array}{ll}
 \frac{2}{\pi\sqrt{1-\tan^2\phi}}\frac{(-\tan \phi)^r}{r}+2^{\frac{5}{2}}\frac{1}{1+\tan \phi}\frac{(-\tan\phi)^{\frac{r}{2}}}{N\sqrt{\pi r}}+\ldots & r\!=\!2m \\
 \frac{2}{\pi\sqrt{1-\tan^2\phi}}\frac{(-\tan \phi)^r}{r}+2^{\frac{3}{2}}\frac{(-\tan\phi)^{\frac{1}{2}}+(-\tan\phi)^{-\frac{1}{2}}}{1+\tan \phi}\frac{(-\tan\phi)^{\frac{r}{2}}}{N\sqrt{\pi r}}+\ldots& r\!=\!2m\!+\!1
\end{array}
\right. \\
\!\!\!\!\!\!\!\!\!\!\!\!\!\!\!\!\!\!\!\!\!\!\!\!\!\!\!\!\!\!\!\!\!\!\!\!
\label{zz-xAFM} \braket{\sigma_j^z \sigma_{j+r}^z} \! &\! 
 \stackrel{r \to \infty}{\simeq} \! & \!
\left\{
\begin{array}{ll}
0  & r\!=\!2m \\
-\frac{2}{\pi} \frac{\tan^r\phi}{r^2}+2^{\frac{3}{2}}\sqrt{1-\tan^2\phi}\frac{(-\tan \phi)^{\frac{r-1}{2}}}{N\sqrt{\pi r}}\ldots & r\!=\!2m\!+\!1
\end{array}
\right.
\end{eqnarray}

While they imply quite clearly that $m_y = m_z =0$ (in accordance to expectations), the extraction of $m_x$ is more subtle: using the standard prescription of taking \mbox{$N \to \infty$} first, one would get $m_x= \left( 1 - \tan^2 \phi \right)^{1/4}$. 
However, one could argue~\cite{Giampaolo18} that a better procedure would be to evaluate eq.~(\ref{xx-xAFM}) at antipodal points $r \sim N/2$ to minimize the correlations and then take the thermodynamic limit. 
In this way, one would get $m_x = \frac{1}{N }\left( 1 - \tan^2 \phi \right)^{1/4} \stackrel{N \to \infty}{\rightarrow} 0$.

\begin{figure}[t]
\begin{center}
    \includegraphics[width=.8\columnwidth]{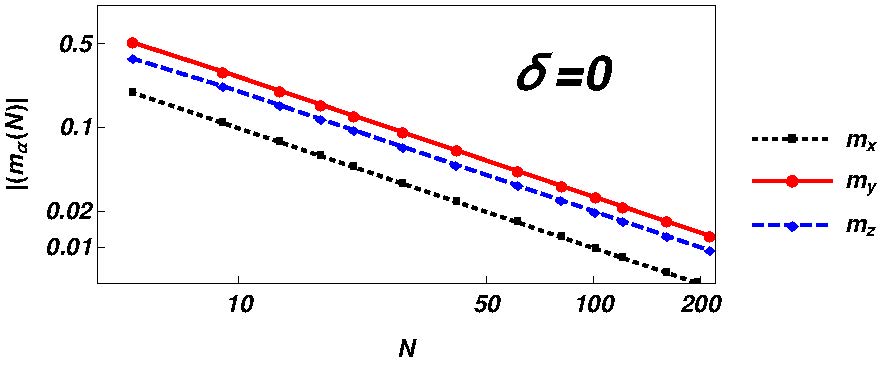}
    \includegraphics[width=.8\columnwidth]{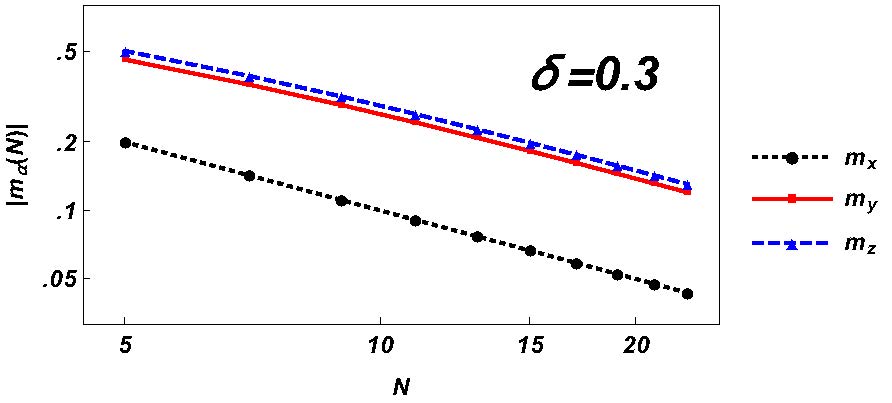} 
\end{center}
    \caption{
    Magnetizations along the three axis (in absolute value) as a function of the chain length for the xAFM/MFM phase at $\phi= -0.25$. 
    The upper panel dots represent the data gathered settings $\delta=0$ using the ``trick'' discussed in the text to evaluate the magnetizations as determinants of $\frac{N-1}{2}\times \frac{N-1}{2}$ matrices, while the dots in the lower one are obtained taking \mbox{$\delta =.3$} and using exact numerical diagonalization. 
    Regardless of the value of $\delta$, we see how all magnetizations decay just algebraically to zero, as shown by the best fit lines (plots presented in log--log--scale).}
    \label{fig:AntiFerroN}
\end{figure}

It is thus important that we can directly access the single spin magnetization using eq.~(\ref{trick}). 
Once more, the expectation values can be cast as determinants of Toeplitz matrices,  whose behaviors are depicted in the upper panel of Fig.~\ref{fig:AntiFerroN}: all magnetizations are characterized by an algebraic decay to zero with the system size. The analytical results for the XY chain are given in the appendix, and demonstrate clearly this property.

Several elements are surprising in these results. 
The most evident one is that FBC kills the magnetization in the $x$--direction, that on an open or even-length chain would be finite, thus seemingly contradicting the independence of Landau construction from boundary conditions. 
Note that a finite spontaneous magnetization can be measured in any finite system, although it decreases algebraically with the system size, a phenomenon we term ``{\it mesoscopic magnetization}''. 
Quite surprisingly, however, this finite--size spontaneous magnetization is {\it not staggered}, but rather ferromagnetic--looking (thus, we will call the AFM phase with FBC, a {\it mesoscopic ferromagnetic} phase, MFM). 
In hindsight, we could have expected this, since a staggered magnetization would have not been compatible with PBC with an odd number of sites (note that this is not a problem for the 2--point function, since its range naturally does not extend beyond one periodicity). 

\begin{figure}[t]
\begin{center}
\includegraphics[width=.68\columnwidth]{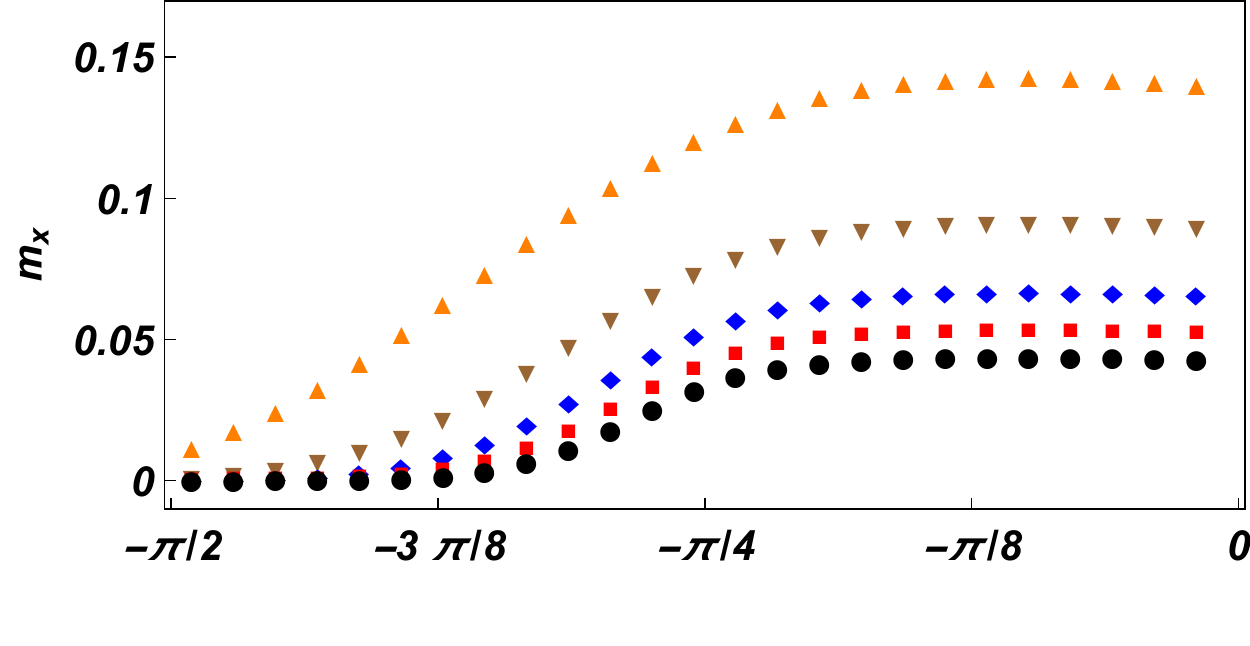}
    \includegraphics[width=.68\columnwidth]{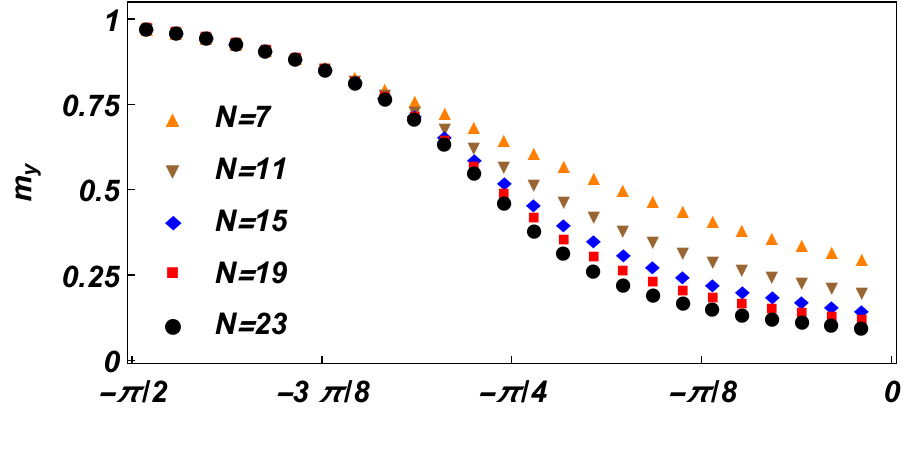}
    \includegraphics[width=.68\columnwidth]{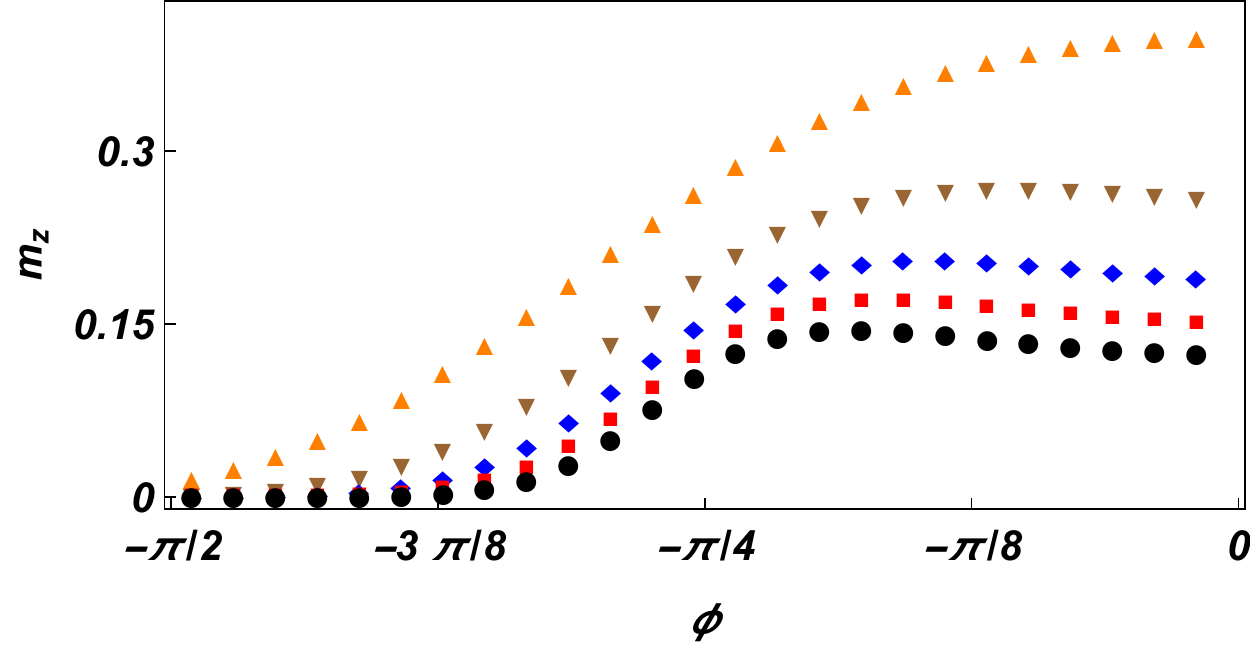} 
\end{center}
    \caption{Plot of the magnetizations as a function of $\phi$ for $\delta = .3$ for several system sizes. The yFM phase ($\phi < -\pi/4$) shows a fast approach to saturation, while for the frustrated case the decay toward zero is algebraically slow.}
    \label{fig:FiniteSizeMag}
\end{figure}

This analytical outcomes are corroborated by exact numerical diagonalization results, which allow us to extended our analysis to the XYZ ($\delta \ne 0$) ring, (see the lower panel of Fig.~\ref{fig:AntiFerroN}).
In Fig.~\ref{fig:FiniteSizeMag} we plot the behavior of the magnetizations as a function of $\phi$ for $\delta =0.3$ for several chain lengths $N$: while in the yFM phase there is little dependence on $N$ as the saturation values are reached quickly, in the MFM phase we observe the slow, algebraic decay toward zero of the order parameters.

It is rather surprising that a finite chain, unable to sustain AFM order, would nonetheless generate a ferromagnetic spontaneous magnetization and that in any finite system, a phase with a dominant interaction along the $x$ direction would show the weakest spontaneous magnetization in that direction, with $m_y$ being the strongest one (once more, these magnetization refers to different states $\ket{g_\alpha}$).
Finally, we remark that FBC also seems to somewhat spoil the cluster decomposition, since the non--staggered mesoscopic magnetization we find is not compatible with (\ref{xx-xAFM}), although both of them vanish in the thermodynamic limit. 

\section{Conclusions} 

We have presented a comparative study of the ferromagnetic and AFM frustrated case for a XYZ chain, showing that, contrary to expectations, the boundary conditions are able to destroy local order. 
We have done so, by realizing that, with no external field, we can exploit particle/hole duality to construct an exact ground state at finite sizes that break the $\mathbb{Z}_2$ symmetry. 
For the XY chain, we can express the one-point function as the determinant of a Toeplitz matrix and evaluate it analytically, while for the interacting cases we can numerically diagonalize the model and calculate the expectation values. 
We benchmarked these procedures on a ferromagnetic phase with FBC to show that they reproduce the expected results eq.~(\ref{yFMOrdParam}), while in an AFM phase the magnetizations, while finite in a finite chain, decay toward zero algebraically in the thermodynamic limit. 
Furthermore, despite a dominant AFM interaction, no magnetization shows a staggered behavior: we thus term this {\it pseudo--phase} generated by FBC a mesoscopic ferromagnetic phase (MFM). 

While we worked at zero fields to have an exact degeneracy on any finite chain and thus to have perfect control of finite size effects, we expect our results to remain valid also with a finite external field (at least up to some threshold). While the exact degeneracy is lifted by a finite field, the energy difference between the lowest energy states in the two parity sectors closes in the thermodynamic limit. However, while this closing is usually exponentially fast, with frustration is only algebraic and accompanied by a similar behavior with respect of other low energy states~\cite{Dong16,Giampaolo18}. This different behavior is mirrored by the two-point function (\ref{xx-xAFM}), which can be extended with a qualitative difference for finite $h$ and indicates a vanishing order parameter when evaluated at antipodal points~\cite{Giampaolo18}. The extension of our analysis to finite $h$ will be the subject of future work, but the existence of a MFM phase would be experimentally detectable, with signatures like those in Fig.~\ref{fig:FiniteSizeMag} easily measurable. The reason for which it has not been seen until now lies in the (surely not extreme) difficulty in realizing a ring geometry and in the expectation that every boundary condition would yield the same result, an expectation that our work put into question.

Nonetheless, we should remark that our results are fully consistent with a straightforward perturbative calculation starting from the classical frustrated Ising chain, as we show in the appendix. Our original contribution is to have found an exact way to approach the thermodynamic limit and to calculate the order parameter. Moreover, with our techniques, we can prove that our result is resilient against a coupling defect breaking translational invariance~\cite{Maric20}, which further ensures the experimental observability of our claim.

Our results are surprising because they seemingly contradict the assumption that boundary conditions cannot influence the bulk behavior of a system and therefore certainly not destroy local order. 
We do not know at the moment how to reconcile this apparent paradox and we invite the community to help us in looking for a general explanation. 
For the moment we can contribute with a couple of observations. 
The first is that FBC provides a non--local contribution to the system since frustration arises from an incompatibility between local and global order. 
Thus, it is possible that the problem we consider can have a topological origin that defies the Landau paradigm. 
Another, somewhat more technical angle, is that in our class of models, the single spin magnetization is dual to a non--local correlator, see eq.~(\ref{trick}). 
From this point of view, it is not surprising that a non--local function is sensitive to the boundary conditions. 
Nonetheless, we have to admit that it seems to us rather paradoxical to consider single--site magnetizations as non--local quantities. 
 
\section*{Acknowledgments}

We acknowledge support from the European Regional Development Fund -- the Competitiveness and Cohesion Operational Programme (KK.01.1.1.06 -- RBI TWIN SIN).
The work is also partially supported by the Croatian Science Fund Project No. IP--2016--6--3347. 
SMG and FF also acknowledge support from the QuantiXLie Center of Excellence, a project co--financed by the Croatian Government and European Union through the European Regional Development Fund -- the Competitiveness and Cohesion
Operational Programme (Grant KK.01.1.1.01.0004) and Croatian Science Fund Project No. IP--2019--4--3321.

\appendix

\section{Supplementary Material}

The Hamiltonian of the topologically frustrated XY chain ($\delta=0$) in a zero magnetic field can be written as
\begin{equation}\label{supp_Hamiltonian_1}
H=\cos\phi \sum_{j=1}^N \sigma_j^x\sigma_{j+1}^x +\sin\phi \sum_{j=1}^N \sigma_j^y\sigma_{j+1}^y \; ,
\end{equation}
where $\phi\in[-\pi/2,0]$ is the parameter that allows us to change the relative weight of the ferromagnetic and antiferromagnetic Hamiltonian terms, $\sigma_j^\alpha$ are the Pauli operators, $N=2M+1$ is the odd number of lattice sites in the system and we assume periodic boundary conditions, i.e. $\sigma_{N+j}^\alpha=\sigma_{j}^\alpha$. 

Due to the absence of any external field, the Hamiltonian in eq.~(\ref{supp_Hamiltonian_1}) commutes with all the three parity operators $\Pi^\alpha=\bigotimes_{j=1}^{N} \sigma_j^\alpha$. 
Because we are considering only systems made by an odd number of spins $N$, such parity operators do not commute with each other. 
Indeed we have
\begin{equation}
\label{supp_parity_relation}
\left[ \Pi^\alpha, \Pi^\beta \right]=\imath\,\varepsilon^{\alpha,\beta,\gamma} 2 (-1)^\frac{N-1}{2} \Pi^\gamma, 
\end{equation}
where $\varepsilon^{\alpha,\beta,\gamma}$ is the Levi--Civita symbol.
The existence of several operators that commute with the Hamiltonian but do not commute with each other induces a degeneracy in all the eigenstates of the Hamiltonian.
We have that to any eigenvalue is associated a $2d$--times degenerate manifold ($d$ positive integer). 
For any of the parity operators $\Pi^\alpha$, each eigenstate manifold will contain $d$ eigenstates belonging to the even sector and $d$ to the odd one.
This is also valid for the ground state manifold which always has a minimum size equal to two.
As we shall see, this property plays a fundamental role in the analytical evaluation of magnetizations in the different spin directions.

\subsection{Solution of the topologically frustrated XY model}

\label{SM_sec_sol}
It is possible to diagonalize analytically the spin model in eq.~(\ref{supp_Hamiltonian_1}) employing the well--known techniques based on a Jordan--Wigner transformation that maps spins into spinless fermions. 
Once we have obtained a spinless fermionic model, a Fourier transformation followed by a Bogoliubov rotation allows us to separate it in $N$ non--interacting fermionic problems that can be analytically treated~\cite{Franchini17}.
At the end of this process, the Hamiltonian in eq.~(\ref{supp_Hamiltonian_1}) can be written as 
\begin{equation}
\label{supp_Hamiltonian_2}
H=\frac{1+\Pi^z}{2}H^+ \frac{1+\Pi^z}{2} + \frac{1-\Pi^z}{2}H^- \frac{1-\Pi^z}{2} \; ,
\end{equation}
where
\begin{equation}
\label{supp_Hamiltonian_3}
H^\pm=\sum\limits_{q\in\Gamma^\pm}^{} \varepsilon(q) \left(a_q^\dagger a_q-\frac{1}{2}\right) ,
\end{equation}
and 
\begin{eqnarray}
\label{supp_energy_momenta}
\epsilon(q) &=&2 \left| \cos\phi \ e^{\imath 2q} + \sin\phi \right| ,\  q\neq 0, \pi \ , \nonumber \\
\epsilon(0) &=&-\epsilon(\pi)=2(\cos\phi+\sin\phi)\; ,
\end{eqnarray}
with the two sets of momenta given, respectively, by $\Gamma^-=\{2\pi k/N \}$ and $\Gamma^+=\{2\pi (k+\frac{1}{2})/N \}$ with $k$ running on all integers between $0$ and $N-1$. 
It is worth to note that the momenta $0\in \Gamma^-$ and $\pi\in\Gamma^+$ (if $N$ is even $\pi\in\Gamma^-$), are different from the others because a) they do not have a corresponding opposite momentum; b) their energies can be negative. 

From eqs.~(\ref{supp_Hamiltonian_2}--\ref{supp_energy_momenta}) it is easy to determine the ground states of the system starting from the vacuum of Bogoliubov fermions in the two sectors ($\ket{0^\pm}$) and taking into account the negative energy modes.
When $-\pi/2<\phi<-\pi/4$ the $0$--mode has negative energy while the $\pi$--mode a positive one.
Therefore, the state with the minimum energy in the odd sector ($a^\dagger_0\ket{0^-}$) has an odd number of fermions while the one in the even sector is characterized by an even number of fermions ($\ket{0^+}$). 
Having, both states the right parity and the same energy they represent a basis for the two-fold degenerate ground states manifold of the Hamiltonian, that is separated, from the rest of the eigenstates, by a finite energy gap that does not close as $N$ increases.

On the contrary, when $-\pi/4<\phi<0$ the energy of the $0$--mode becomes negative while the one of the $\pi$--mode becomes positive.
As a consequence, the state with the minimum energy in the even sector ($a^\dagger_\pi\ket{0^+}$) has now an odd number of fermions while the one in the odd sector ($\ket{0^-}$) has an even one. 
Therefore they cannot represent physical states of our system, as they violate the parity constraint of their relative sectors. 
On the contrary, the ground states can be recovered from such states with the minimum energy by adding the lightest possible excitation. 
Because we are considering $\phi\in[-\pi/2,0]$, the smallest excitations are associated with the $\pi$ and the $0$--mode respectively for the even and the odd sector.
Therefore in the region $\phi\in[-\pi/2,0]$ the two ground states of the Hamiltonian that are also eigenstates of $\Pi^z$ are 
\begin{eqnarray}
\label{supp_groundstate}
\ket{g^+}=\ket{0^+} & \;\;\;\;\; &  \mathrm{even\; sector} \nonumber \\
\ket{g^-}=a^\dagger_0\ket{0^-} & \;\;\;\;\; &  \mathrm{odd \;sector} 
\end{eqnarray}
Note that, although the expressions in eq.~(\ref{supp_groundstate}) describe the ground states in both the xFM and xAFM phases, they in fact characterize quite different structures. 
For instance, in the frustrated case, since the GS is obtained as the lightest excitation on top of the lowest possible energy state (as just explained), adding different excitations provides states with an almost continuum of energy, which becomes a dense, gapless band in the thermodynamic limit~\cite{Dong16,Dong17}.

Moreover, also the different correlations and magnetizations in the two phases show different behaviors.
This can be traced back to the fact that, in the two phases, the Majorana correlation functions are different.
Indeed, defining the Majorana fermionic operators as 
\begin{equation}
\label{supp_Majorana_operators}
A_j \equiv \left(\bigotimes_{l=1}^{j-1} \sigma_l^z \right) \sigma_j^x \; , \quad B_j \equiv \left(\bigotimes_{l=1}^{j-1} \sigma_l^z \right) \sigma_j^y  \ ,
\end{equation}
and exploiting Wick's Theorem, all non--vanishing spin correlation functions on ground states that are also eigenstates of $\Pi^z$ can be evaluated in terms of the two--body Majorana correlation functions:
\begin{eqnarray}
\label{supp_Majorana_correlators}
\bra{g^\pm}A_{j+r}A_j\ket{g^\pm} &= &
\bra{g^\pm}B_{j+r}B_j\ket{g^\pm}=\delta_{r0} \\
\bra{g^\pm}A_{j+r}B_j\ket{g^\pm}&=&\frac{\imath}{N}\sum\limits_{q\in\Gamma^\pm}e^{i2\theta_{q}}e^{-iqr}+\frac{2 \imath }{N} f^\pm(r)\nonumber\,.
\end{eqnarray}
In eqs.~(\ref{supp_Majorana_correlators}) $\theta_{q}$ stands for the Bogoliubov angle satisfying
\begin{equation} 
\label{supp_Bogoljubov_angles}
e^{\imath 2\theta_{q}}=e^{\imath q} \frac{\cos\phi+\sin\phi \ e^{-\imath 2q}}{|\cos\phi+\sin\phi \ e^{-\imath 2q}|} \; ,
\end{equation}
while the function $f^\pm(r)$ is zero for $-\pi/2<\phi<-\pi/4$, while for $-\pi/4<\phi<0$ we have $f^+(r)=(-1)^{r}$ and $f^-(r)=-1$.

\subsection{Two--spins correlation function along the $x$ and $y$ directions:}

Let us now move to analyze the behavior of the two--spin correlation functions along $x$ and $y$ directions as a function of $r$. 
Following the path traced in Ref.~\cite{Barouch71,LSM-1961}, it is easy to express such correlations in terms of determinants of a $r\times r$ Toeplitz matrices.
More precisely said $C_{xx}(r)$ the two--spin correlation function along $x$ at distance $r$, i.e. $C_{xx}(r)=\bra{g^\pm} \sigma^x_j \sigma^x_{j+r} \ket{g^\pm}$, we have that it is given by
\begin{equation}
\label{corr_xx_1}
C_{xx}(r)=(-\imath)^r \Delta(\rho_{xx})
\end{equation}
where $\Delta(\rho_{xx})$ is the determinant of the matrix $\rho_{xx}$
\begin{equation}
\!\!\!\!\! \label{corr_xx_2}
\rho_{xx}\!\equiv\!\left(\!\!
\begin{array}{ccccc}
G(1)\! & G(0) \!&  G(-1) \!&  \!\cdots\! & G(2-r)\! \\
G(2)\! & G(1)\! & G(0)\! & \!\cdots\! & G(3-r)\! \\
G(3)\! & G(2) \!& G(1) \!& \!\cdots\! & G(4-r) \\
\vdots & \vdots  & \vdots  & & \vdots \\
G(r)\! & G(r\!-\!1)\! & G(r\!-\!2)\! \!& \!\cdots\! & G(1)\!
\end{array}
\!\!\right) \,,
\end{equation}
where $G(r)\equiv -\imath \bra{g^+}A_{j+r}B_j\ket{g^+}$.
At the same time the correlation function along $y$ at distance $r$ is given by
\begin{equation}
\label{corr_yy_1}
C_{yy}(r)=\bra{g^\pm} \sigma^y_j \sigma^y_{j+r} \ket{g^\pm}=(\imath)^r \Delta(\rho_{yy})
\end{equation}
where $\Delta(\rho_{yy})$ is the determinant of the matrix $\rho_{yy}$
\begin{equation}
\!\!\!\!\! \label{corr_yy_2}
\rho_{yy}\!\equiv\!\left(\!\!
\begin{array}{ccccc}
G(-1)\! & G(0) \!&  G(1) \!&  \!\cdots\! & G(r-2)\! \\
G(-2)\! & G(-1)\! & G(0)\! & \!\cdots\! & G(r-3)\! \\
G(-3)\! & G(-2) \!& G(-1) \!& \!\cdots\! & G(r-4) \\
\vdots & \vdots  & \vdots  & & \vdots \\
G(-r)\! & G(1\!-\!r)\! & G(2\!-\!r)\! \!& \!\cdots\! & G(-1)\!
\end{array}
\!\!\right) 
\end{equation}

As we anticipated, the behavior of these two correlation functions is very different in the two phases of our system.
In the ferromagnetic phase, the asymptotic behavior is well known from the literature: $C_{xx}(r)$ exponentially decays to zero, while $C_{yy}(r)$ saturates exponentially fast to the square of the $y$-magnetization, that is to $\sqrt{1-\cot^2 \phi}$.

In the xAFM phase, the evaluation of the asymptotic behaviors of the Toeplitz determinants is more complicated, but can also be done analytically. The details of the analytical computation are given in \cite{Maric-math}, where such determinants are studied. The results are given in the main text, and in this work we checked them numerically.

\subsection{Magnetizations along the $x$ and $y$ directions}

In this section, we show how it is possible to exploit the particular symmetries of the model in eq.~(\ref{supp_Hamiltonian_1}), to evaluate, for any odd $N$, the magnetization along the $x$ and the $y$ directions. 
For sake of simplicity, we limit ourselves to illustrate the method for the magnetization along the $x$ direction and we report the results for both at the end. 

As we have seen, in the region that we are analyzing, the ground state manifold has always dimension equal to two.
Therefore the set made by $\ket{g^+}$ and $\ket{g^-}$ represents a good basis for the ground state manifold and, hence, all its elements can be written as a linear combination of $\ket{g^+}$ and $\ket{g^-}$. 
But in our case, we can say more. 
As we have already shown, the Hamiltonian in eq.~(\ref{supp_Hamiltonian_1}) commutes not only with $\Pi^z$ but also with $\Pi^x$ and $\Pi^y$.
This fact implies that the state $\Pi^x \ket{g^+}$ is also a ground state of the system.  
On the other hand, taking into account the anticommutation rules of the spin operators on the same site and the fact that we are considering a system with odd $N$, it is easy to see that, while $\ket{g^+}$ is in the even sector of $\Pi^z$, $\Pi^x \ket{g^+}$ lives in the odd one.
As a consequence we have that $\ket{g^-}=\Pi^x \ket{g^+}$, up to a global multiplicative phase factor, and hence the generic ground state can be written as
\begin{equation}
\label{supp_generic_groundstate_x}
\ket{g}=
\Big[
\cos(\theta)+ \sin(\theta) e^{\imath \psi} \Pi^x
\Big] \ket{g^+} \: ,
\end{equation}
or equivalently using $\Pi^y$, since $\ket{g^-}$ and $\Pi^y \ket{g^+}$ only differ by a global phase factor.

Let us now choose a generic site $j$ of the system.
For the generic ground state in eq.~(\ref{supp_generic_groundstate_x}) the magnetization along $x$ on the $j$-th spin is 
\begin{eqnarray}
\label{supp_mag_x_1}
\! m_x(j)&\! =\! & \bra{g} \sigma^x_j \ket{g}  \\
\! & \! = \! & \cos^2(\theta) \bra{g^+} \sigma^x_j \ket{g^+} +\sin^2(\theta) \bra{g^+} \Pi^x \sigma^x_j \Pi^x \ket{g^+} \nonumber \\
\! &\!+\! & \frac{1}{2}\sin(2\theta)\!
\Big[ e^{\imath \psi} \! \bra{g^+}\! \sigma^x_j \Pi^x\! \ket{g^+}\!+\!
e^{-\imath \psi} \! \bra{g^+}\! \Pi^x \sigma^x_j  \!\ket{g^+}
\Big] \nonumber 
\end{eqnarray}
Being both $\ket{g^+}$ and $\Pi^x \ket{g^+}$ eigenstates of $\Pi^z$, the two expectation values $\bra{g^+} \sigma^x_j \ket{g^+}$ and $\bra{g^+} \Pi^x \sigma^x_j \Pi^x \ket{g^+} $ vanish.
On the contrary, because the number of spins in the system is odd, the operator $\Pi^x \sigma^x_j=\sigma^x_j \Pi^x$, that is equal to $\tilde{\Pi}_j^x=\bigotimes_{l\neq j} \sigma^x_l$, is an operator that commutes with $\Pi^z$ and hence can have a non--vanishing expectation value on $\ket{g^+}$. 
Therefore we have
\begin{eqnarray}
\label{supp_mag_x_2}
\! m_x(j)&\! =\! & \cos(\psi)\sin(2\theta)\!
\bra{g^+}\! \tilde{\Pi}_j^x \ket{g^+}  \, . 
\end{eqnarray}
which reaches the maximum for $\psi=0$ and $\theta=\frac{\pi}{4}$, that is, the state on which we focus in the letter.

Hence, to evaluate the magnetization, we only need to determine  the expectation value $\bra{g^+}\! \tilde{\Pi}_j^x \ket{g^+} $. 
Since $[\tilde{\Pi}_j^x,\Pi^z]=0$, the magnetization can be easily evaluated exploiting the representation of $\tilde{\Pi}_j^x$ in terms of the Majorana operators in eq.~(\ref{supp_Majorana_operators}) and Wick's theorem.
Without loss of generality, let us set $j=1$. 
From the definition of the Majorana operators in eq.~(\ref{supp_Majorana_operators}), the operator $\tilde{\Pi}_j^x$ can be written as
\begin{equation}
\label{supp_step_1}
\tilde{\Pi}_1^x =(-\imath)^{\frac{N-1}{2}} \bigotimes_{l=1}^{\frac{N-1}{2}} B_{2l}A_{2l+1} \,, 
\end{equation}
and, exploiting Wick's theorem, we obtain that the expectation value $\bra{g^+}\! \tilde{\Pi}_1^x \ket{g^+} $ is
\begin{equation}
\label{supp_step_1_bis}
\bra{g^+}\! \tilde{\Pi}_1^x \ket{g^+} =(-1)^{\frac{N-1}{2}} \Delta(\rho_x) \,, 
\end{equation}
where $\Delta(\rho_x)$ is the determinant of the $\frac{N-1}{2}\times\frac{N-1}{2}$ Toeplitz matrix $\rho_x$ that reads  
\begin{equation}
\!\!\!\!\! \label{supp_step_2}
\rho_x\!=\!\left(\!\!
\begin{array}{ccccc}
G(1)\! & G(-1) \!&  G(-3) \!&  \!\cdots\! & G(4\!-\!N)\! \\
G(3)\! & G(1)\! & G(-1)\! & \!\cdots\! & G(6\!-\!N)\! \\
G(5)\! & G(3) \!& G(1) \!& \!\cdots\! & G(8\!-\!N) \\
\vdots & \vdots  & \vdots  & & \vdots \\
G(N\!-\!2)\! & G(N\!-\!4)\! & G(N\!-\!6)\! \!& \!\cdots\! & G(1)\!
\end{array}
\!\!\right) 
\end{equation}
with $G(r)\equiv -\imath \bra{g^+}A_{j+r}B_j\ket{g^+}$

On the other hand, the magnetization along $y$ on the spin $1$ becomes
\begin{eqnarray}
\label{supp_mag_y_2}
\! m_y(1)&\! =\! & \cos(\psi)\sin(2\theta)\!
\bra{g^+}\! \tilde{\Pi}_1^y \ket{g^+}  \, . 
\end{eqnarray}
where $\tilde{\Pi}_1^y=\bigotimes_{l=2}^N \sigma^y_l$. 
Also in this case the maximum of the magnetization is equal to $\bra{g^+}\! \tilde{\Pi}_1^y \ket{g^+}$, which in turn can be written as 
\begin{equation}
\label{supp_step_1_bis_y}
\bra{g^+}\! \tilde{\Pi}_1^y \ket{g^+} =(-1)^{\frac{N-1}{2}}  \Delta(\rho_y) \,, 
\end{equation}
where $\Delta(\rho_y)$ is the determinant of the $\frac{N-1}{2}\times\frac{N-1}{2}$ Toeplitz matrix $\rho_y$ 
\begin{equation}
\!\!\!\!\! \label{supp_step_2_y}
\rho_y\!=\!\left(\!\!
\begin{array}{ccccc}
G(-1)\! & G(-3) \!&  G(-5) \!&  \!\cdots\! & G(2\!-\!N)\! \\
G(1)\! & G(-1)\! & G(-3)\! & \!\cdots\! & G(4\!-\!N)\! \\
G(3)\! & G(1) \!& G(-1) \!& \!\cdots\! & G(6\!-\!N) \\
\vdots & \vdots  & \vdots  & & \vdots \\
G(N\!-\!4)\! & G(N\!-\!6)\! & G(N\!-\!6)\! \!& \!\cdots\! & G(-1)\!
\end{array}
\!\!\right) 
\end{equation}

\subsection{Magnetizations in the ferromagnetic phase}

If we are in the yFM phase the expressions for the magnetizations in the thermodynamic limit can be obtained analytically in the following way. We have that in eq.~(\ref{supp_Majorana_correlators}) the function $f^+(r)=0$ and hence $G(r)$ becomes 
	\begin{equation}
	\label{supp_step_3}
	G(r)=\frac{1}{N}\sum_{q\in\Gamma^+} \frac{\cos\phi+\sin\phi \ e^{-\imath 2q}}{|\cos\phi+\sin\phi \ e^{-\imath 2q}|} e^{-\imath q(r-1)} \,,
	\end{equation}
	and for large $N$ we can approximate the sum with an integral, hence obtaining
	\begin{equation}
	\label{supp_step_4}
	G(r)=\frac{1}{2\pi}\int_0^{2\pi} \frac{\cos\phi+\sin\phi \ e^{-\imath 2q}}{|\cos\phi+\sin\phi \ e^{-\imath 2q}|} e^{-\imath q(r-1)} dq \,.
	\end{equation}
	
	To evaluate the determinants  $\Delta(\rho_{x,y})$ of the Toeplitz matrices in eq.~(\ref{supp_step_2}) eq.~(\ref{supp_step_2_y}) we introduce 
	\begin{equation}
	D_n \equiv G(2n-1)=-\frac{1}{2\pi}\int_0^{2\pi} \frac{1+\cot\phi \ e^{\imath q}}{|1+\cot\phi \ e^{\imath q}|} e^{-\imath qn} dq \,,
	\label{symboltrick}
	\end{equation}
	and rewrite them as
	\begin{equation}
	\Delta_r (\rho_x)=\begin{vmatrix}
	D_1 & D_{0}& \dots & D_{2-r} \\
	D_2 & D_{1}& \dots & D_{3-r} \\
	\vdots & \vdots& \ddots & \vdots\\ 
	D_{r} & D_{r-1}& \dots & D_{1} \\
	\end{vmatrix} , \quad r=\frac{N-1}{2} \; ,
	\label{rhoxtrick}
	\end{equation}
	and
	\begin{equation}
	\Delta_r (\rho_y)=\begin{vmatrix}
	D_0 & D_{-1}& \dots & D_{1-r} \\
	D_1 & D_{0}& \dots & D_{2-r} \\
	\vdots & \vdots& \ddots & \vdots\\ 
	D_{r-1} & D_{r-2}& \dots & D_{0} \\
	\end{vmatrix} , \quad r=\frac{N-1}{2} \; .
	\label{rhoytrick}
	\end{equation}
	
	The latter can be evaluated straightforwardly for large $N$ using Szeg\"{o} theorem~\cite{Hirschman}, yielding, to leading order, 
	\begin{equation}
	\label{supp_step_5}
	\bra{g^+} \tilde{\Pi}_1^y \ket{g^+}=(1-\cot^2\phi)^\frac{1}{4} .
	\end{equation}
	
	The magnetization in the $x$ direction, instead, is more complicated, because the generating function of the corresponding Toeplitz matrix has a non--zero winding number. 
	To overcome this problem, we proceed as in Ref.~\cite{Wu66} and notice that the determinant in eq.~(\ref{rhoxtrick}) can be seen as the minor of $\Delta_{r+1} (\rho_y)$ in eq.~(\ref{rhoytrick}) obtained removing the first row and the last column. 
	To calculate this minor, we use Cramer's rule and consider the following linear problem:
	\begin{equation}
	\sum_{m=0}^{r} D_{n-m} \: x_m = \delta_{n,0} \, , \qquad n=0,\ldots,r \;.
	\end{equation}
	Then, 
	\begin{equation}
	\Delta_r (\rho_x) = (-1)^r x_r \Delta_{r+1} (\rho_y) \: ,
	\label{deltarx}
	\end{equation}
	where $\Delta_{r+1} (\rho_y)$ is a Toeplitz determinant satisfying the conditions for Szeg\"{o} theorem.
	For large $r$, $x_r$ can be evaluated following the standard Wiener--Hopf procedure as in Ref.~\cite{Wu66}. 
	The result is
	\bea
	x_r & \stackrel{r \gg 1}{\sim} &
	- \frac{1}{2 \pi \imath} \oint 
	\frac{\xi^{r-1} \ d \xi }{\sqrt{(1+\cot \phi \ \xi) (1+\cot \phi \ \xi^{-1})}}
	\nonumber \\
	&=&
	- \frac{1}{\pi} \int_{0}^{-\cot \phi}
	\frac{x^{r-1/2} \ d x }{\sqrt{(1+\cot \phi \ x) (-\cot \phi - x)}} \; , 
	\label{xrint}
	\eea
	where we deformed the contour of integration around the branch cut. 
	Up to now, everything has been similar to the standard calculations usually performed in the XY model, but now we have to proceed anew because, unlike the generating functions for the two--point correlators which have two pairs of poles and zero when extended to the complex plane, the generating function (symbol) in eq.~(\ref{symboltrick}) we have for the magnetization only has one movable pole/zero.
	
	Fortunately, the integral in eq.~(\ref{xrint}) can be expressed in terms of hypergeometric functions:
	\bea 
	x_r & \stackrel{r \gg 1}{\sim} &
	- \frac{(-1)^r}{\sqrt{\pi}} \ \frac{\cot^r \phi}{\sqrt{1-\cot^2 \phi}} \ \Gamma \left( r + \frac{1}{2}\right) /  \Gamma \left( r + 1 \right) \times
	\nonumber \\
	&& \times _2 F_1 \left( \frac{1}{2} , \frac{1}{2}; r+1; \frac{\cot^2 \phi}{\cot^2 \phi - 1 }\right)\; , 
	\eea
	whose asymptotic behavior gives to leading order
	\be
	x_r \stackrel{r \gg 1}{\sim}
	- \frac{(-1)^r}{\sqrt{ \pi r}} \ \frac{\cot^r \phi}{\sqrt{1-\cot^2 \phi}} \; , 
	\label{xras}
	\ee
	since the $_2 F_1$ tends to $1$ for large $r$.
	Combining eq.~(\ref{xras}) with eq.~(\ref{deltarx}) and eq.~(\ref{supp_step_5}) we arrive at
	\be
	\Delta_r (\rho_x) = \frac{(-1)^r\cot^r \phi}{(1-\cot^2\phi)^\frac{1}{4}\sqrt{ \pi r}} \  \: ,
	\ee
	which means that the magnetization in the $x$ direction decays exponentially with the system size: 
	\bea
	\label{supp_step_5x}
	\bra{g^+} \tilde{\Pi}_1^x \ket{g^+} & \stackrel{N \gg 1}{\sim} &
	\frac{\cot^{\frac{N-1}{2}} \phi}{(1-\cot^2\phi)^\frac{1}{4}\sqrt{ \pi (N-1)/2}} 
	\nonumber \\
	& \stackrel{N \to \infty }{\sim} & 0 \: .
	\eea
	Note that, despite the $x$ interaction being AFM, the corresponding magnetization is not staggered.
	
	Finally, the magnetization in the $z$--direction is just equal to the Majorana two--point function in eq.~(\ref{supp_Majorana_correlators}) and thus its exponential decay to zero arises as to the difference between the finite sum in eq.~(\ref{supp_Majorana_correlators}) and vanishing of the corresponding integral in the $N \to \infty$ limit.
	
	\subsection{Magnetizations in the frustrated phase}
	
	If we are in the xAFM phase we have that in eq.~(\ref{supp_Majorana_correlators}) the function $f^+(r)\neq0$ and hence the generating function $G(r)$ becomes 
	\begin{equation}
	\label{supp_step_3_bis}
	G(r)\!=\frac{2}{N}(-1)^r+\!\frac{1}{N}\!\sum_{q\in\Gamma^+}\frac{\cos\phi+\sin\phi \ e^{-\imath 2q}}{|\cos\phi+\sin\phi \ e^{-\imath 2q}|} e^{-\imath q(r-1)} \,,
	\end{equation}
	in which, for large $N$, we can approximate the sum with an integral, hence obtaining
	\begin{equation}
	\label{supp_step_6}
	G(r)=\frac{2}{N}(-1)^r+
	\frac{1}{2\pi}\int_0^{2\pi}  \frac{\cos\phi+\sin\phi \ e^{-\imath 2q}}{|\cos\phi+\sin\phi \ e^{-\imath 2q}|} e^{-\imath q(r-1)} dq \,.
	\end{equation}
	This generating function reflects the fact that, effectively, the ground states of the frustrated case have a single, delocalized excitation.
	Thus, in this phase, we write the Toeplitz determinants in eq.~(\ref{supp_step_2}) and eq.~(\ref{supp_step_2_y}) as
	\begin{equation}
	\Delta_r (\rho_x)=\begin{vmatrix}
	\tilde{D}_0 & \tilde{D}_{-1}& \dots & \tilde{D}_{1-r} \\
	\tilde{D}_1 & \tilde{D}_{0}& \dots & \tilde{D}_{2-r} \\
	\vdots & \vdots& \ddots & \vdots\\ 
	\tilde{D}_{r-1} & \tilde{D}_{r-2}& \dots & \tilde{D}_{0} \\
	\end{vmatrix} , \quad r=\frac{N-1}{2} \; ,
	\label{rhoxtrick1}
	\end{equation}
	and
	\begin{equation}
	\Delta_r (\rho_y)=\begin{vmatrix}
	\tilde{D}_{-1} & \tilde{D}_{-2}& \dots & \tilde{D}_{-r} \\
	\tilde{D}_0 & \tilde{D}_{-1}& \dots & \tilde{D}_{1-r} \\
	\vdots & \vdots& \ddots & \vdots\\ 
	\tilde{D}_{r-2} & \tilde{D}_{r-3}& \dots & \tilde{D}_{-1} \\
	\end{vmatrix} , \quad r=\frac{N-1}{2} \; ,
	\label{rhoytrick1}
	\end{equation}
	where 
	\begin{equation}
	\tilde{D}_n \equiv G(2n+1)=
	-\frac{2}{N}(-1)^n+\frac{1}{2\pi}\int_0^{2\pi} D 
	\left( e^{\imath q} \right)  e^{-\imath qn} dq  ,
	\label{tildeDdef}
	\end{equation}
	with
	\be
	D \left( e^{\imath q} \right) \equiv 
	\frac{1+\tan\phi \ e^{- \imath q}}{|1+\tan\phi \ e^{-\imath q}|} \: .
	\label{DxAFM}
	\ee
	Note that, compared to the definitions employed for the yFM phase, we changed the definition of the generating function by shifting its Fourier series, so that eq.~(\ref{DxAFM}) has zero winding number.
	Using these determinant representations the magnetizations can be computed analytically, similarly to the two-point correlation functions. The details of the analytical computation are given in \cite{Maric-math}. The result, for large (odd) $N$, is
		\begin{align}
		&\bra{g^+} \tilde{\Pi}_1^x \ket{g^+}=(-1)^{\frac{N-1}{2}}\frac{1}{N}(1-\tan^2\phi)^{\frac{1}{4}},\\
		&\bra{g^+} \tilde{\Pi}_1^y \ket{g^+}=\frac{2}{N}\frac{(1-\tan\phi)^{\frac{1}{4}}}{(1+\tan \phi)^{\frac{3}{4}}}.
		\end{align}	
	Here we check the results numerically, as shown in the main text.
	
	\paragraph{Two--spins correlation function and magnetization along the $z$ direction in the frustrated phase}
		The correlation function along $z$ is simply the determinant $\Delta(\rho_{zz})$ of the matrix
		\begin{equation}
		\rho_{zz}\equiv \begin{pmatrix}
		G(0) & G(-r) \\
		G(r) & G(0)
		\end{pmatrix}.
		\end{equation}
	The integral in \eqref{supp_step_6} can be studied by deforming the contour around the branch cuts and using the properties of hypergeometric functions, as in \cite{McCoy68,Barouch71,Wu66}. In this way we get
	\begin{eqnarray}
	\!\!\!\!\!\!\!\!\!\!\!\!\!\!\!\!\!\!\!\!
	G(r)&=&\frac{2}{N}(-1)^r, \qquad\qquad\qquad\qquad\qquad\qquad\qquad\qquad r=2m \; ; \\
	\!\!\!\!\!\!\!\!\!\!\!\!\!\!\!\!\!\!\!\!
	G(r)&\stackrel{r\to\infty}{\simeq}&\frac{2}{N}(-1)^r+\sqrt{2(1-\tan^2\phi)}\frac{(-\tan\phi)^{\frac{r-1}{2}}}{\sqrt{\pi r}}\;\qquad\quad\;\; r=2m+1 \; ;\\
	\!\!\!\!\!\!\!\!\!\!\!\!\!\!\!\!\!\!\!\!
	G(-r)&\stackrel{r\to\infty}{\simeq}&
	\frac{2}{N}(-1)^r -\sqrt{\frac{2}{1-\tan^2\phi}}\frac{(-\tan\phi)^{\frac{r+1}{2}}}{\sqrt{\pi r^3}} \qquad\qquad\quad\! r=2m+1 \; .
	\end{eqnarray}
	Using these formulas we get the result \eqref{zz-xAFM} in the main text. Finally, the magnetization in the $z$ direction is given by $G(0)$ so it is simply equal to $ \frac{2}{N}$.

	\subsection{Perturbative analysis of the frustrated phase}
	
A further insight into low-energy behavior of the model in the frustrated phase and a further check on the results on magnetization can be obtained through a perturbative analysis close to the classical Ising point $\phi=0$. This analysis is similar to the one of Ref.~\cite{Campostrini2015b}. At the classical point it is trivial to diagonalize the model, in the basis where $\sigma_j^x$ are diagonal. The ground states are simply the kink states $\ket{j}$ and $\Pi^z\ket{j}$, for $j=1,2,...,N$, where the state
\begin{equation}
\ket{j}=\ket{...,1,-1,1,1,-1,1,....} 
\end{equation}
has a ferromagnetic bond between the sites $j$ and $j+1$, with $\sigma_j^x=\sigma_{j+1}^x=1$, and antiferromagnetic bonds between other adjacent sites, while the state $\Pi^z\ket{j}$, with all spins reversed, has  $\sigma_j^x=\sigma_{j+1}^x=-1$ and all the other bonds antiferromagnetic. The ground state energy equals $-N+2$. The states $\ket{j}$ and $\Pi^z\ket{j}$ belong to the parity sectors $\Pi^x=(-1)^{(N-1)/2}$ and $\Pi^x=-(-1)^{(N-1)/2}$ respectively.

For $\phi<0$, the $\sigma_j^y \sigma_{j+1}^y$ terms kick in, splitting the $2N$-fold ground state degeneracy. The corresponding eigenstates and the correction to the energies are found by diagonalizing the perturbation in the ground state subspace (other states are separated by a finite gap and thus can be neglected at this level). Since the matrix elements of the perturbation between two different $\Pi^x$ sectors vanish (because the $\sigma_j^y \sigma_{j+1}^y$ terms still commute with all the parities $\Pi^\alpha$), we can focus on each sector separately. In the $\Pi^x=(-1)^{(N-1)/2}$ sector they read
\begin{equation}
\bra{l} \sum_{j=1}^N \sigma_j^y\sigma_{j+1}^y \ket{k} =\delta_{l,k-2}^{(N)}+\delta_{l,k+2}^{(N)} \; ,
\end{equation}
where the equalities in the Kronecker delta $\delta^{(N)}$ are understood modulo $N$. It follows that the perturbation in the subspace spanned by $\ket{j}$, $j=1,...,N$, is a cyclic matrix
	\begin{equation}
	\sum_{j=1}^N \sigma_j^y\sigma_{j+1}^y =
	\begin{pmatrix}
	c_0 & c_{N-1} & \dots & c_2 & c_1\\
	c_1 & c_{0} & \dots & c_3 & c_2\\
	\vdots & \vdots & \ddots & \vdots & \vdots \\
	c_{N-2} & c_{N-1} & \dots & c_0 & c_{N-1}\\
	c_{N-1} & c_{N-2} & \dots & c_1 & c_0\\
	\end{pmatrix} , 
	\end{equation}
	with
	\begin{equation}
	c_j=\delta_{j,2}+\delta_{j,N-2}
	\end{equation}
	Diagonalizing the cyclic matrix~\cite{McCoyWu} we find the energies
	\begin{equation}
	E_q=-(N-2)\cos\phi+2\sin\phi \cos (2q) , \quad q\in\Gamma^- ,
	\end{equation}
	corresponding to the states
	\begin{equation}\label{supp_superposition_kinks}
	\ket{s_q}=\frac{1}{\sqrt{N}} \sum_{j=1}^{N}e^{iq}\ket{j}
	\end{equation}
	Clearly, the states with opposite $\Pi^x$ corresponding to the same energies can be constructed as $\Pi^z\ket{s_q}$. The energies of the exact solution reduce, of course, to those of perturbative calculation when $\phi$ is close to $0$.
	
	The energy is minimized by $q=0$ and thus the ground states are translationally invariant superpositions of kinks, $\ket{s_{q=0}}=\sum_{j=1}^{N}\ket{j}/\sqrt{N}$ and $\Pi^z\ket{s_{q=0}}$. By looking at the parities we can conclude that the states eq.~\eqref{supp_groundstate} from the exact solution, in the limit $\phi\to0^-$, are equal (up to a phase factor) to
	\begin{eqnarray}\label{supp_identification_of_ground_states}
	\ket{g^+}&=\frac{1}{\sqrt{2}} (1+\Pi^z)\ket{s_{q=0}}\\
	\ket{g^-}&=\frac{1}{\sqrt{2}} (1-\Pi^z)\ket{s_{q=0}}
	\end{eqnarray}
	Having identified these ground states, we can compute analytically the magnetization and the two-point correlator in the generic ground state eq.~\eqref{supp_generic_groundstate_x}. For this task we use the relation
	\begin{equation}\label{supp_kink_states_magnetization}
	\bra{l} \sigma_j^x \ket{l}=
	\begin{cases}
	(-1)^{l+j+1} , & l<j\\
	(-1)^{l+j} , & l\geq j
	\end{cases}
	\end{equation}
	that follows from the definition of the kink states.
	
	The two-point correlator $C_{xx}(r)=\bra{g}\sigma_j^x\sigma_{j+r}^x\ket{g}$ has the same value on whole ground state subspace and is, in particular, equal to
	\begin{equation}
	\bra{s_{q=0}}\sigma_j^x\sigma_{j+r}^x\ket{s_{q=0}}=\frac{1}{N} \sum_{l=1}^{N} \bra{l}\sigma_{j}^x \ket{l} \bra{l} \sigma_{j+r}^x\ket{l}  .
	\end{equation}
	Using eq.~\eqref{supp_kink_states_magnetization} and summing up we find
	\begin{equation}
	C_{xx}(r)=(-1)^r\bigg(1-\frac{2r}{N}\bigg),
	\end{equation}
	in agreement with the result obtained from the exact solution.
	
	The magnetization is determined by the element 
	\begin{equation}
	\bra{s_{q=0}}\sigma_j^x\ket{s_{q=0}}=\frac{1}{N}\sum_{l=1}^{N} \bra{l} \sigma_j^x \ket{l} =\frac{1}{N}.
	\end{equation}
	Using its value it follows that in the generic ground state eq.~\eqref{supp_generic_groundstate_x} the magnetization is equal to
	\begin{equation}
	m_x(j) = (-1)^{(N-1)/2} \cos(\psi)\sin(2\theta) \frac{1}{N}
	\end{equation}
	The magnetization is ferromagnetic and its value agrees with the one obtained from the exact solution.

\vskip .5cm

\hrulefill


\vskip .5cm


\begin{thebibliography}{99}


\bibitem{LandauStatPhys}
   L.D. Landau, E.M. Lifshitz, \& L.P. Pitaevskij, {\it Statistical Physics}, Pergamon Press, Oxford (1978).
   
\bibitem{ChandlerBook}
  D. Chandler, {\it Introduction to Modern Statistical Mechanics}, Oxford University Press; 1 edition (1987).   
   
\bibitem{AndersonNotions}
  P.W. Anderson, {\it Basic Notions Of Condensed Matter Physics}, Addison-Wesley (1997).
  
\bibitem{ColemanBook}
  P. Coleman, {\it Introduction to Many-Body Physics}, Cambridge University Press (2016).
  
\bibitem{StoneBook}
  M. Stone (Ed.), {\it Quantum Hall Effect}, World Scientific (1992).

\bibitem{WenBook}
  X.-G. Wen, {\it Quantum Field Theory of Many-body Systems: From the Origin of Sound to an Origin of Light and Electrons}, Oxford University Press (2004).
  
\bibitem{Nayak2008}
C. Nayak, S.H. Simon, A. Stern, M. Freedman, \& S. Das Sarma, Rev. Mod. Phys. {\bf 80}, 1083 (2008).
\\ {\it Non-abelian anyons and topological quantum computation}.   

\bibitem{Hasan2010}
  M.Z. Hasan \& C.L. Kane, Rev. Mod. Phys. {\bf 82}, 3045 (2010)
\\ {\it Colloquium: topological insulators}.  
  
\bibitem{FradkinBook}
  E. Fradkin, {\it Field theories of condensed matter physics}, Cambridge University Press (2013).
  
\bibitem{BernevigBook}
B. A. Bernevig \& T.L. Hughes, {\it Topological Insulators And Topological Superconductors}, Princeton University Press (2013).

\bibitem{Giampaolo15}
S. M. Giampaolo \& B. C. Hiesmayr, Phys. Rev. A {\bf 92}, 012306 (2015). 
\\{\it Topological and nematic ordered phases in many--body cluster--Ising models}

\bibitem{Witten16}
  E. Witten, Rev. Mod. Phys. {`bf 88}, 35001 (2016).
\\ {\it Free fermions And Topological Phases}.

\bibitem{ZengBook}
   B. Zeng, X. Chen, D.-L. Zhou, \& X.-G. Wen, {\it Quantum Information Meets Quantum Matter: From Quantum Entanglement to Topological Phases of Many-Body Systems}, Springer (2019).

\bibitem{Dong16}
  J.-J. Dong, P. Li, \& Q.-H. Chen, J. Stat. Mech. P113102 (2016)
\\ {\it The A-Cycle Problem for Transverse Ising Ring.}

\bibitem{Giampaolo18}
S. M. Giampaolo, F. B. Ramos, \& F. Franchini, J. Phys. Commun. {\bf 3}, 081001 (2019)
\\ {\it The Frustration in being Odd: Area Law Violation in Local Systems.}

\bibitem{Affleck95}
   I. Affleck, Acta Phys. Polon {\bf B 26}, 1869 (1995)
\\ {\it Conformal Field Theory Approach to the Kondo Effect}.

\bibitem{Durganandini96}
   P. Durganandini, Phys. Rev. {\bf B 53}, R8832(R) (1996).
\\ {\it Kondo effect in a Luttinger liquid: A boundary-conformal-field-theory approach}.

\bibitem{Cardy84}
   J. Cardy, Nucl. Phys. {\bf B 240}, 4 (1984)
\\ {\it Conformal invariance and surface critical behavior}.

\bibitem{Cardy04}
   J. Cardy, arXiv:hep-th/0411189 (2004).
\\ {\it Boundary Conformal Field Theory}.

\bibitem{DiFrancescoBook}
    P. Di Francesco, P. Mathieu, \& D. Senechal, {\it Conformal Field Theory}, Springer (1999).
    
\bibitem{KorepinBook}
   V.E. Korepin, N.M. Bogoliubov, \& A.G. Izergin, {\it Quantum Inverse Scattering Method and Correlation Functions}, Cambridge University Press (1997).

\bibitem{Korepin00}
  V.E. Korepin \& P. Zinn-Justin,  J. Phys. {\bf A 33}, 7053 (2000)
\\ {\it Thermodynamic limit of the six-vertex model with domain	wall boundary conditions}.

\bibitem{ZinnJustin02}
  P. Zinn-Justin,  (2002), arXiv:cond-mat/0205192 (2002).
\\ {\it The influence of boundary conditions in the six-vertex model}.

\bibitem{Colomo10}
  F. Colomo \& A. G. Pronko, J. Stat. Phys. {\bf 138}, 662 (2010)
\\ {\it The arctic curve of the domain-wall six-vertex model}.

\bibitem{BleherBook}
 P. Bleher \& K. Liechty, {\it Random Matrices and the Six-Vertex Model}, CRM monographs series, vol. 32, American Mathematical Society, Providence (2013).
 
\bibitem{Colomo16}
   F. Colomo \& A. Sportiello, J. Stat. Phys. {\bf 164}, 1488 (2016)
\\ {\it Arctic curves of the six-vertex model on generic domains: the Tangent Method}.
 
\bibitem{Allegra16}
  N. Allegra, J. Dubail, J.-M. St\'ephan, \& J. Viti, J. Stat. Mech. 2016, 053108 (2016).
\\ {\it Inhomogeneous field theory inside the arctic circle}.

\bibitem{Reshetikhin17}
  N. Reshetikhin \& A. Sridhar, Commun. Math. Phys. {\bf 356}, 535 (2017)
\\ {\it Integrability of limit shapes of the six-vertex model}.

\bibitem{DiFrancesco18}
   P. Di Francesco \& E. Guitter, J. Phys. A: Math. Theor. {\bf 51}, 355201 (2018)
\\ {\it Arctic curves for paths with arbitrary starting points: a tangent method approach}.

\bibitem{Colomo18}
   F. Colomo, A.G. Pronko, \& A. Sportiello, J. Stat. Phys. {\bf 174}, 1 (2018)
\\ {\it Arctic Curve of the Free-Fermion Six-Vertex Model in an L-Shaped Domain}.

\bibitem{Toulouse77}
  G. Toulouse, Commun. Phys. {\bf 2}, 115 (1977)
\\ {\it Theory of the frustration effect in spin glasses: I}. 

\bibitem{Vannimenus77}
 J. Vannimenus \& G. Toulouse, J. Phys. C {\bf 10}, L537 (1977)
\\ {\it Theory of the frustration effect. II. Ising spins on a square lattice}.

\bibitem{Wolf03}
   M.M. Wolf, F. Verstraete \& J.I. Cirac, Int. Journal of Quantum Information {\bf 1}, 465 (2003)
\\ {\it Entanglement and Frustration in Ordered Systems}.

\bibitem{Giampaolo11}
   S. M. Giampaolo,  G. Gualdi, A. Monras, \& F. Illuminati, Phys. Rev. Lett. {\bf 107}, 260602 (2011)
\\ {\it Characterizing and quantifying frustration in quantum many-body systems}.

\bibitem{Marzolino13}
  U. Marzolino, S. M. Giampaolo, \& F. Illuminati, Phys. Rev. A {\bf 88}, 020301(R) (2013)
\\ {\it Frustration, entanglement, and correlations in quantum many body systems}.

\bibitem{Sadoc07}
  J. F. Sadoc \& R. Mosseri, {\it Geometrical frustration}. Cambridge University Press (2007).
  
\bibitem{Lacroix11}
  C. Lacroix, P. Mendels, \& F. Mila (eds), 
{\it Introduction to Frustrated Magnetism: Materials, Experiments, Theory}.  
Springer Series in Solid-State Sciences, Vol. 164 (2011). 

\bibitem{Diep13}
  H. T.  Diep, {\it Frustrated Spin Systems}, World Scientific (2013).

\bibitem{Wannier50}
  G.H. Wannier, Phys. Rev. {\bf 79}, 357 (1950)
\\ {\it Antiferromagnetism. The Triangular Ising Net}.

\bibitem{Burkhards1985}
    T.W. Burkhardt \& I. Guim, J. Phys. {\bf A}: Math. Gen {\bf 18}, L33 (1985)
\\ {\it Finite-size scaling of the quantum Ising chain with periodic, free, and antiperiodic boundary conditions.}

 \bibitem{Cabrera1987}
   G.G. Cabrera \& R. Jullien, Phys. Rev. {\bf B 35}, 7062 (1987)
\\ {\it Role of boundary conditions in the finite-size Ising model.}

\bibitem{Campostrini15}
   M. Campostrini, A. Pelissetto, \& E. Vicari, Phys. Rev. {\bf E 91}, 042123 (2015)
\\ {\it Quantum transitions driven by one-bond defects in quantum Ising rings.}

\bibitem{Ercolessi13}
  E. Ercolessi, S. Evangelisti, F. Franchini, \& F. Ravanini, Phys. Rev. B 88, 104418 (2013)
\\ {\it Modular invariance in the gapped XYZ spin-$\frac{1}{2}$ chain.}

\bibitem{Franchini17}
F. Franchini, {\it An introduction to integrable techniques for one-dimensional quantum systems}, Lecture Notes in Physics {\bf 940}, Springer (2017).

\bibitem{Maric19-1}
V. Mari\'{c}, S. M. Giampaolo, D. Kui\'{c}, \& F. Franchini, In preparation
\\ {\it The Frustration in being Odd: Exact finite size degeneracies.}

\bibitem{Maric19-2}
V. Mari\'{c}, S. M. Giampaolo, D. Kui\'{c}, \& F. Franchini, In preparation
\\ {\it The Frustration in being Odd: Can Boundary Conditions induce a Quantum Phase Transition?.}

\bibitem{Sachdev11}
S. Sachdev,  {\it Quantum Phase Transitions}, Cambridge University Press (2011).

\bibitem{damski12}
  B. Damski \& M. M. Rams, J. Phys. {\bf A 47}, 025303 (2014)
\\{ \it Exact results for fidelity susceptibility of the quantum Ising model: The interplay between parity, system size, and magnetic field.}

\bibitem{LSM-1961}
E. Lieb, T. Schultz, \& D. Mattis, Ann. of Phys. {\bf 16}, 407-466
(1961)
\\ {\it Two Soluble Models of an Antiferromagnetic Chain.}

\bibitem{Jordan28}
P. Jordan \& E. Wigner, Z. Phys. {\bf 47}, 631 (1928)
\\ {\it \"Uber das Paulische \"Aquivalenzverbot}.
	
\bibitem{McCoy68}
B.M. McCoy, Phys. Rev. {\bf 173}, 531 (1968)
\\ {\it Spin Correlation Functions of the X-Y Model.}
	
\bibitem{Barouch71}
E. Barouch \&  B.M. McCoy, Phys. Rev. {\bf A 3}, 786 (1971)
\\ {\it Statistical Mechanics of the XY Model. II. Spin-Correlation	Functions.}
	
\bibitem{Dong17}
J.-J. Dong \& P. Li, Mod. Phys. Lett. {\bf B 31}, 1750061 (2017)
\\ {\it The a-cycle problem in XY model with ring frustration}.
	
\bibitem{Maric-math}
V. Mari\'{c}, \& F. Franchini, 
\\ {\it Asymptotic behavior of Toeplitz determinants with delta function singularities}, arXiv:2006.01922 (2020).  

\bibitem{Maric20}
V. Mari\'{c}, S. M. Giampaolo, \& F. Franchini, In preparation
\\ {\it The Frustration in being Odd: Resilience against perturbations.}

\bibitem{Hirschman}
I.I. Hirschman, Jr., Amer. J. Math. {\bf 88}, 577 (1966).
\\ {\it The Strong Szeg\"o Limit Theorem for Toeplitz Determinants.}

\bibitem{Wu66}
T.T. Wu, Phys. Rev. {\bf 149}, 380 (1966).
\\ {\it Theory of Toeplitz Determinants and the Spin Correlations of the Two-Dimensional Ising Model. I.}

\bibitem{Campostrini2015b}
M. Campostrini, A. Pelissetto, \& E. Vicari, J. Stat. Mech. P11015 (2015)
\\ {\it Quantum Ising chains with boundary fields.}

\bibitem{McCoyWu}
B. McCoy \& T.T. Wu, {\it The Two-Dimensional Ising Model}, Harvard University Press (1973).



 

\end{thebibliography}
\end{document}